\documentclass[11pt]{article}

\usepackage[utf8]{inputenc}
\usepackage[margin=1in]{geometry}
\usepackage[hidelinks]{hyperref}
\usepackage{url}
\title{Selling the Stock, Not the Cream:\\
The Soviet \'Emigr\'e Career Premium of the 1990s}
\author{Vitaly Pronskikh\\[2pt]
\small Center for Slavic, Eurasian, and East European Studies (CSEEES),\\
\small University of North Carolina at Chapel Hill, Chapel Hill, NC, USA}
\date{July 2026}

\begin{document}
\maketitle

\begin{abstract}
In the early-to-mid 1990s, scientists emigrating from the former Soviet Union to the United States---especially physicists, engineers, chemists, and biologists---frequently secured prestigious and highly visible academic positions, including professorships, named chairs, and laboratory leadership roles; scientists of comparable ability who arrived after about 2000 generally built more modest, less visible, and often non-academic careers. Rejecting the common view that this
reflects the people---the elite having left first---we set aside the thin apex of
Nobel- and Fields-level émigrés and study the much larger cohort of capable but
non-stellar scientists, showing that the same kind of scientist fared differently
according to the year of arrival; the explanation therefore lies in the structure
of the receiving market, not in individual ability. Interpreting these appointments as merely the natural reward of exceptional individual brilliance risks survivorship bias, since later Nobel-level recognition makes visible only the most successful cases and obscures the broader demand for Soviet scientific capital. We weigh four conditions that
were favourable in the 1990s and closed by the mid-2000s: the technology transfer
and export of a finite, distinctive stock of Soviet expertise that commanded a
career premium; the favourable immigration regime created by the Soviet Scientists
Immigration Act of 1992; the surge of U.S.-trained Chinese and Indian competitors;
and the securitizing aftermath of 11~September 2001. All four were materially
in play, with technology transfer and export primary: its premium opened the
window, and its depletion---as the exported knowledge was published and absorbed
into global science---removed the demand on which the other factors depended. A
further, cross-cutting mechanism, the cultural ``ghettoization'' of émigrés into
co-national laboratory enclaves, capped their scientific visibility and independent
advancement. I argue that the imbalance between the émigré generations was structural, not
personal.
\end{abstract}

\section{Introduction: the puzzle and the thesis}
 
The disintegration of the Soviet Union in 1991 precipitated one of the largest
scientific migrations of the modern era. The collapse of research funding and the
disruption of an entire national research system drove a massive outflow of
scientists to the United States, Western Europe, and, later,
Asia~\cite{shiltsev-diaspora,schweitzer}. Estimates of its scale depend on
definition---bibliometric reconstructions from publication records place the number
of emigrating researchers over the 1990s--2010s on the order of fifteen thousand,
with perhaps two to three thousand in particle and nuclear physics alone, equivalent
to some five to seven per cent of the world workforce in those
fields~\cite{shiltsev-diaspora,ganguli}---but by any measure the exodus was
comparable in magnitude to the celebrated flight of European scientists to America in
the 1930s and 1940s~\cite{shiltsev-diaspora}. It was concentrated in the natural and
technical sciences and peaked in the 1990s and the very early 2000s~\cite{scopus}.
This paper concerns a puzzle about the careers of those who left. Scientists who
arrived in the United States in the early and middle 1990s frequently secured
prestigious and visible positions---tenured professorships, named chairs, and the
leadership of laboratory groups. Those who arrived after about 2000, by contrast,
built markedly more modest, less visible, and more often non-academic careers. Why?

The prevailing answer is one of individual quality. In the dominant account of the
post-Soviet scientific exodus---developed most fully by Graham and
Dezhina~\cite{graham-dezhina}---the earliest émigrés were disproportionately the
scientific elite: senior, internationally known figures whose reputations preceded
them and who therefore commanded the best positions, while later departures drew
increasingly on the rank and file. A widely held perception sharpens this into a
simple contrast: the first wave was outstanding and the rest were ordinary, and the
difference in their fortunes merely reflects the difference in their gifts---the cream
left first. On this reading the career gap requires no explanation beyond the changing
composition of the flow.

Two further alternatives should be granted explicitly. First, part of the post-2000
shift away from academia may reflect choice rather than exclusion: the dot-com and
biotechnology booms drew scientists into industry and start-ups, so that some absence
from the professoriate represents an attractive outside option taken rather than an
academic door closed. Second, the end of the Cold War may have lowered the strategic
salience of Soviet science---in contrast to earlier, security-driven brain
gains---and with it the special institutional willingness to accommodate FSU
scientists; this consideration is not wholly separate from our first factor, since the
premium we describe was partly strategic in origin. Neither alternative overturns the
structural account, but both qualify it, and a fuller treatment would weigh
career-destination data (academia versus industry) for the two cohorts directly.
 
We take a different view, and a different unit of analysis. It is not in dispute that
the most celebrated figures---scientists of Nobel and Fields calibre---left early and
rose brilliantly; but that thin apex is not the interesting case, and to explain its
success by its brilliance is very nearly tautological. The cohort that matters for the
puzzle is the far larger body of capable but non-stellar scientists---competent
researchers of ordinary distinction---who nonetheless secured genuine academic careers
in the 1990s and who could not have done so a decade later. For this cohort the
quality explanation fails a basic test: a scientist of the same calibre fared very
differently according to the year in which he happened to arrive. That pattern points
away from the attributes of the people and toward the structure of the market that
received them; the simple equation of the first wave with excellence and the later
flow with mediocrity has, in any case, not gone unchallenged~\cite{myths-reality}.
Our argument is accordingly a structural and demand-side one. The career gap, we
contend, was produced not by a decline in the abilities of successive émigrés but by a
configuration of external conditions that was unusually favourable in the 1990s and
that closed, more or less together, by the middle of the 2000s.

A word on scope, terms, and evidence. By ``former-Soviet (FSU) scientists'' we mean
research scientists and engineers in science-and-engineering (S\&E) fields who were
trained in, or emigrated from, the Soviet Union and its successor states; we use this
term consistently in preference to narrower labels. Our explanandum is academic and
research-career attainment---placement in tenured or tenure-track university
positions, the leadership of laboratory groups, and comparable independent
standing---rather than elite distinction alone; and the contrast we draw runs between
those who entered in the first post-collapse wave (roughly 1992--1996) and those who
entered after about 2000, observed into the mid-2000s. The evidence is of two kinds.
For aggregate scale and timing we draw on bibliometric reconstructions of the émigré
population~\cite{ganguli,scopus} and on official statistics that permit breakdowns by
country of citizenship or birth---the NSF Survey of Earned Doctorates~\cite{nsf-sed-citizenship},
the \emph{Science and Engineering Indicators}~\cite{nsf-sei}, and the Department of
Homeland Security immigration yearbooks~\cite{dhs-yearbook}. For the texture of
careers we rely on named cases and institutional records, and, for the enclave mechanism of Section~\ref{sec:ghetto-labs}, the author's own participant observation, which illustrate mechanisms
that the aggregates cannot. We state the principal limitation at the outset: there is
no comprehensive registry of FSU émigré scientists linked to career outcomes, so the
argument combines population-level aggregates with illustrative cases rather than
resting on a single outcome dataset.

Our account is primarily demand-side, but supply dynamics modulate it and deserve
notice. The first wave was driven out by the economic collapse of 1991--1993, an
indiscriminate exodus that placed a large number of scientists on the market at once;
that this glut nonetheless produced strong career outcomes is itself evidence of how
powerful the contemporaneous demand was, since abundant supply was absorbed on
favourable terms. The pattern reversed in the 2000s: as the Russian economy recovered,
the outflow shrank to a small fraction of the research workforce~\cite{intellectual-potential,scopus},
and those who still left were, if anything, a more self-selected group. This cuts in
favour of our thesis rather than against it: a later cohort that was, on average,
\emph{more} select yet fared \emph{worse} is poorly explained by any decline in
individual quality and well explained by the closing of the favourable conditions we
examine.
 
We identify four such conditions and devote a section to each. The first is the
transfer and export of unique Soviet scientific and technological capital by the first cohorts: a stock
of distinctive expertise, accumulated under decades of comparatively isolated
development, that commanded a premium in the American research market but that
was---precisely because it was a finite stock rather than a renewable flow---eventually
depleted as it was extracted, published, and absorbed into global science. The second
is U.S.\ immigration policy: the Soviet Scientists Immigration Act of 1992, which
classified former-Soviet scientists as persons of ``exceptional ability'' and waived
the job-offer requirement, lowering the friction of entry so that the premium could be
converted quickly into positions---an instrument that lapsed in 1996 and, after a brief
revival, expired in 2006. The third is competition: the drastic enlargement, by the
2000s, of the Chinese and Indian doctoral inflow, a sustained and self-renewing
pipeline of U.S.-trained scientists who competed for the same positions and who, having
been formed inside the American system, held a structural advantage in its norms and
networks. The fourth is the aftermath of 11~September 2001, whose consequences for
science policy, for the perception of foreigners, and hence for the informal networks
on which academic careers depend, formed a broad headwind for the foreign-born at
large. We will argue that, taken together, these four factors bore far more weight on
the careers of the émigré cohort \emph{en masse} than did any superior individual
ability of those who came first.
 
\section{Why Soviet scientific capital was unique---and field-specific}

Soviet science developed in substantial autonomy from the West, with limited exchange of people, journals, and ideas. The cost of that isolation was real---lag in several fields---but it had a reverse side: in certain disciplines the Soviet schools followed fundamentally different trajectories and produced original methods, instruments, and theoretical frameworks that simply did not exist in the West. It was this \emph{divergence of trajectories}, rather than abstract ``quality,'' that constituted the market value of the transferred capital.

Crucially, that value was unevenly distributed. It was concentrated in physics (especially theoretical, nuclear, high-energy, and plasma physics), in mathematics, in materials science and metallurgy, and in selected branches of chemistry and applied mechanics---fields in which Soviet schools were both strong and unlike their Western counterparts. In areas where Soviet science lagged, the unique transfer was correspondingly thinner. This field-specificity must be kept in mind: the ``abundance'' of the 1990s was an abundance in particular niches, not a uniform front.

Two illustrations fix the order of magnitude. In materials research, Ukrainian and Russian expertise in superhard coatings and powder metallurgy was of keen interest to U.S. researchers, including the military, and advanced Soviet work on nanomaterials fed successful commercial ventures~\cite{sciencediplomacy}. In experimental physics, scientists at the Soviet weapons laboratory VNIIEF produced, in the mid-1990s, a world-record magnetic field on the order of 28~million gauss---about 50~million times the field at the Earth's surface---in joint high-field experiments regarded as among the best instrumented of their kind~\cite{cisac}. Results of this character were, in effect, a currency: a demonstration of a level of capability for access to which Western institutions were prepared to open doors.

\section{The currency of transfer, I: university chairs}

The clearest evidence that unique expertise was paid for in academic opportunity is the roster of \'emigr\'es who were granted senior professorships---frequently named, endowed chairs---at leading U.S. universities within a year or two of departure.

Consider mathematics, where the Moscow and Leningrad schools were world-leading and methodologically distinctive. Grigory (Gregory) Margulis, a Fields Medalist (1978) who would later add the Wolf Prize (2005) and the Abel Prize (2020), emigrated in 1990 and joined the Yale faculty in 1991, where he holds the Erastus~L.~DeForest Professorship of Mathematics~\cite{margulis-yale}. His doctoral adviser at Moscow State University, Yakov Sinai---himself a student of Kolmogorov, and later an Abel laureate (2014)---has been a professor of mathematics at Princeton since 1993, serving as the Thomas Jones Professor in 1997--98~\cite{sinai-princeton}. In each case a U.S. department extended a chair to a mathematician who brought an entire research tradition that was thinly represented locally; no laboratory or equipment was required, so the premium attached purely to the knowledge and the person.

Physics provides a parallel. The cosmologist Andrei Linde, a principal author of inflationary theory, moved from CERN to the United States in 1990 and became a professor of physics at Stanford (today the Harald Trap Friis Professor); his wife, the supergravity theorist Renata Kallosh, took a Stanford professorship in the same move~\cite{linde-kavli}. The pattern---an immediate senior appointment, often a dual-career placement, for carriers of a distinctive theoretical school---recurs across the strong Soviet fields and is the apex form of the ``payment in opportunity'' that defines the 1990s window.

\section{The currency of transfer, II: laboratory leadership}

Where the \'emigr\'es entered national laboratories rather than universities, the same premium took the form of distinguished-scientist appointments and leadership of research groups---again granted as the price of absorbing scarce expertise.

The canonical case is Alexei Abrikosov. The originator of the theory of type-II superconductors---work for which he shared the 2003 Nobel Prize in Physics---Abrikosov emigrated in 1991, as the Soviet Union dissolved, and joined Argonne National Laboratory as a Distinguished Scientist, becoming leader of Argonne's condensed-matter theory group from 1992 to 2000~\cite{abrikosov-anl,abrikosov-nobel}. The Royal Society's biographical memoir is explicit that securing him ``was a real coup for Argonne'' and required deliberate effort by laboratory management---language that captures precisely the competitive value placed on such a recruit~\cite{abrikosov-rs}. (He concurrently held adjunct professorships at the University of Illinois at Chicago and the University of Utah~\cite{abrikosov-nobel}.)

A second case shows the same dynamic for a younger \'emigr\'e who rose into leadership inside the U.S. system. The accelerator physicist Vladimir Shiltsev, trained at the Budker Institute in Novosibirsk, joined Fermilab in 1996 as a Robert~R.~Wilson Fellow, became a Distinguished Scientist, headed the Tevatron Department from 2001 to 2005, and served as the inaugural Director of the Fermilab Accelerator Physics Center from 2007 to 2018~\cite{shiltsev-ae}. Shiltsev is also the author of a systematic study of the post-Cold-War diaspora of Soviet particle physicists, which finds that many \'emigr\'e researchers ``assumed leadership positions, drove major experimental and theoretical initiatives, and achieved scientific distinction equal to or exceeding'' their Soviet-era work---a generalization for which his own trajectory is a representative instance~\cite{shiltsev-diaspora}.

These named figures are, of course, the elite apex of the phenomenon. The analytically decisive point, developed in Section~7, is that the same demand that handed chairs and laboratory leadership to the apex also opened ordinary tenure-track lines to a much broader population of capable but non-stellar \'emigr\'es---and it is the \emph{breadth} of that demand, not merely its apex, that distinguishes the 1990s from the 2000s.

\section{The internal stratification of the \'emigr\'e labour market:
``ghetto laboratories''}

\label{sec:ghetto-labs}
A complementary mechanism, documented by the sociologist Izabela Wagner, helps
explain what happened \emph{beneath} this apex---and why the breadth of demand
did not translate into a corresponding breadth of independent careers. Studying
mobile scientists in the contemporary laboratory world, Wagner identifies a
recurrent formation she calls the ``ghetto laboratory'': a research group led by
an immigrant principal investigator who staffs it largely with compatriots
recruited into subordinate roles---postdoctoral fellows and research associates
rather than future independent investigators~\cite{wagner}. Her larger argument is
that such arrangements expose a gap between the official discourse of
science---open access, democratic and meritocratic selection---and an unspoken
reality in which ethnicity and national origin structure who occupies which rung
of the laboratory hierarchy~\cite{wagner}.

Applied to the former-Soviet cohort, this formation is precisely the texture of
the ``breadth'' invoked above. The same surge of demand that delivered chairs and
laboratory directorships to the most eminent \'emigr\'es also drew a far larger
number of FSU scientists into the United States---but frequently into the
dependent positions inside laboratories headed by their more successful
countrymen, rather than into independent faculty lines of their own. The dynamic
is genuinely double-edged. On one side it offered a landing place: a
Russian-speaking, Soviet-trained newcomer could enter a laboratory where the
scientific culture, language, and tacit standards were already familiar, which
lowered the transaction costs of entry and provided immediate scientific
employment; on the same logic it functioned as a vehicle of transfer in its own
right, since a Soviet-trained principal investigator effectively reproduced the
methods and ethos of his home school by populating the bench with people formed
in the same tradition. On the other side, as Wagner's choice of the word
``ghetto'' signals, the same structure could confine those beneath the principal
to a subordinate, non-independent standing, with restricted scientific visibility
and partial separation from the mainstream networks through which independent
scientific careers are in practice built~\cite{wagner}.

A concrete instance illustrates the pattern---and, importantly, illustrates that
the binding constraint was \emph{structural} rather than personal.~\footnote{The qualitative dimension of this account, and the enclave interpretation placed on
it, are the author's own, drawn from several years of direct participant observation
within a setting of this kind---first-hand, longitudinal evidence the documentary
record cannot supply, resting necessarily on a single site.} Consider a Monte Carlo code for
high-energy radiation transport---the simulation of hadronic and electromagnetic
cascades in particle accelerators, detectors, and shielding, central to the
design and radiation safety of high-energy-physics experiments. Begun in the
mid-1970s at a leading Soviet high-energy-physics institute, the code had, across
the decades, a single coordinating author. In the early 1990s he accepted a
permanent senior appointment at a major U.S.\ high-energy-physics laboratory,
bringing the code with him to continue its development for the laboratory's
experiments and for U.S.\ high-energy-physics needs, where it became a controlled,
registered-distribution government asset. Around this capability the laboratory
established a dedicated group, most of whose former-Soviet, Russian-speaking
members were recruited around or after~2000. For these scientists the group was a genuine
and valuable route into the laboratory; and---a point that matters for what
follows---the group's head appears to have made considerable efforts to advance
his members' standing and to promote their careers.

Yet on the analysis advanced here, the members' prospects were shaped less by
anything their head did or failed to do than by the structural arrangements of the
laboratory in the post-2000 period---arrangements that, in Wagner's terms, made
the group operate as a ``ghetto laboratory''~\cite{wagner} \emph{despite}, rather
than because of, its leadership. Two disclosed features of that structure support
the reading. The first is career classification. Although the members in time held
continuing (non-term) appointments and were, for the most part, naturalized
U.S.\ citizens who underwent the same extensive external review as their
colleagues, they were placed, under the laboratory's own personnel taxonomy, not
in its independent scientific track---the track of the group's head and of most
U.S.-origin scientists who had risen through postdoctoral positions, and the one
the laboratory itself treated as equivalent to a university professorship---but in
a parallel technical track, akin to that of applied physicists or engineers, which
carried little of the independence, professional-service role, or scientific
visibility that the senior scientific track conferred.

The second feature is
integration: the capability was concentrated in a single group whose members
shared a former-Soviet background, and that group remained weakly connected to the
laboratory's other, predominantly U.S.-origin scientific groups. The cumulative
effect was an enclave---a cohort of able, securely employed, regularly reviewed
specialists whose individual visibility and prospects for an independent
scientific career nonetheless remained limited, not through the intent of a
gatekeeper but through an institutional configuration that placed them in the
non-independent track and left them weakly tied to the rest of the laboratory.  Where those conditions obtain,
enclaves arise whatever the goodwill of those who lead them.

One likely reason the group head obtained a senior permanent appointment, substantial autonomy, and authority to recruit and shape the group, while later arrivals remained in subordinate, non-independent positions, is that he brought with him a distinctive and controlled technical asset: the Monte Carlo radiation-transport code of which he had been the primary long-term developer and custodian. This export-controlled capability, which the laboratory adopted as a registered government asset, constituted a concrete form of the unique Soviet scientific capital discussed earlier in this paper. In exchange for transferring and maintaining control over this asset, the institution was prepared to grant him a privileged structural position and the latitude to build a research group around it. Subordinates recruited later, however scientifically skilled, did not arrive with comparable controlled assets and therefore entered an already consolidated enclave on markedly less favourable terms.

This internal stratification carries a temporal payoff for the central puzzle of
the reversal. In the early 1990s the ghetto laboratories were still
\emph{forming}, as the apex scientists arrived and built their groups; the period
therefore generated genuine independent openings as well as subordinate ones. By
the 2000s the configuration had matured---as the case above shows---so that the
opening most readily available to a newly arriving FSU scientist was increasingly
a subordinate, non-independent appointment within an established co-national
laboratory, rather than an independent line of his own. Two consequences compound
the other factors treated in this study. First, placement in a non-independent
track, together with weak integration into the laboratory's other groups, tended
to foreclose the socialization into mainstream American scientific networks on
which an independent career depends, weakening these scientists precisely where
the cohorts trained from the graduate level inside those same
networks were strongest. Second, the very availability of co-ethnic subordinate
labour allowed the system to absorb later \'emigr\'es \emph{as labour} without
advancing them to independent standing, which is consistent with the observed
shift from professorial appointments in the 1990s to more modest, dependent, and
often non-academic careers thereafter. The evidence here is qualitative and the
interpretation of ``ghetto'' is contested---the same structures can be read as
rational co-ethnic adaptation rather than segregation---but as a mechanism it
links the technology-transfer account directly to the downgrading of \'emigr\'e
careers over time.

\section{Case study: the TOPAZ-II space-reactor program}

The most literal example of paying with opportunity for transferred technology---combining hardware, knowledge, and direct employment---is the TOPAZ International Program. TOPAZ-II was a Soviet single-cell, in-core \emph{thermionic} space nuclear power system, developed at the Kurchatov Institute of Atomic Energy and built by the enterprise Krasnaya Zvezda (whose director, Georgiy Gryaznov, led the Soviet delegation), embodying more than two decades of Soviet space-reactor experience maturing between roughly 1969 and 1990~\cite{topaz-casestudy}. Technically it was a compact reactor of about 115~kWt with thermionic fuel elements around each of its fuel pins, NaK liquid-metal cooling, and a zirconium-hydride moderator~\cite{topaz-whatisnuclear}; it was subsequently purchased by the United States expressly for technology transfer and testing~\cite{topaz-inis}.

The contact began at the Albuquerque Space Nuclear Power Symposium, where Academician Nikolay N.~Ponomarev-Stepnoi of the Kurchatov Institute first described the TOPAZ-II reactor and signalled interest in a partnership---an episode dated by the principal program history to January 1989.\footnote{The literature is not fully consistent on the dating of the initial contact. The principal case study places Ponomarev-Stepnoi's first description of TOPAZ-II at Albuquerque in January 1989~\cite{topaz-casestudy}; other sources emphasize the January 1991 (8th) symposium, at which a full TOPAZ-II model was exhibited.} On the U.S.\ side, several engineers recognized the commercial and strategic potential of the technology and obtained funding from the SDIO to arrange delivery and testing in Albuquerque~\cite{topaz-casestudy}. The reactors were shipped to the New Mexico Engineering Research Institute at the University of New Mexico for non-nuclear ground testing supported by Phillips Laboratory, Sandia National Laboratories, and Los Alamos National Laboratory~\cite{topaz-inis,topaz-whatisnuclear}; the SDIO ultimately purchased reactors for roughly \$13~million.

For the present argument the decisive feature is employment. The program ``presented the only opportunity for Russian scientists and engineers to continue the development'' of thermionic space power for civil applications once Soviet funding had been cut off, and it ``eventually employ[ed] up to fifty Russian scientists, engineers, and technicians'' in the United States~\cite{topaz-casestudy}. Here the abstract claim of this chapter becomes concrete: a unique Soviet technology was transferred as both apparatus and tacit know-how, and the carriers of that know-how were paid in continued scientific employment that their home system could no longer provide.

\section{The currency of transfer, III: the rank and file}
 
The figures named in the last two sections are the apex of the phenomenon, and
they are its most visible signature; but the analytically central claim of this
paper concerns not them but the far larger population beneath them, and the
evidence for that breadth is statistical rather than anecdotal. To explain the
success of a Fields Medalist by his brilliance is, as noted at the outset, very
nearly tautological; what the quality account cannot accommodate is the ordinary
researcher of no special celebrity who nonetheless secured a genuine academic
post in the 1990s and could not have done so a decade later. The same sources
used above to gauge the scale of the exodus also gauge its absorption. Some ten
thousand FSU scientists and engineers entered the United States across the
1990s~\cite{ganguli}, and a large share of the early wave---on the order of
seventy per cent---found positions in universities and research-and-development
organizations~\cite{ird-funding}: a placement rate that an ordinary academic
market does not extend to a mid-career foreign cohort, and that is intelligible
only as the footprint of exceptional demand operating well below the level of the
celebrated few.
 
Mathematics shows the breadth most
precisely. Of approximately one thousand mathematicians who left the former
Soviet Union in the early 1990s, about a third settled in the United
States~\cite{borjas-doran}; and in the single academic year 1991--92, scientists
from Eastern Europe and the Soviet Union accounted for some fifteen per cent of
all tenured and tenure-track appointments in U.S.\ mathematics
departments~\cite{borjas-doran}---the Soviet-trained being much the larger part of
that share. The overwhelming majority of these new hires were not medalists but
ordinary research mathematicians whose schools and methods happened to be scarce
in the West. That a single mid-career foreign nationality could supply, in a given
year, roughly one in seven of an entire discipline's permanent appointments is the
breadth of demand made quantitative; the aforementioned two eminent chairs were
merely its most conspicuous instances.
 
The same structure is visible in experimental physics, where the aforementioned apex case sat atop a numerous rank and file. Bibliometric reconstruction places
two to three thousand former-Soviet specialists in particle and nuclear physics in
the post-1991 diaspora~\cite{shiltsev-diaspora}, and they were absorbed not as
celebrated leaders but as staff scientists, postdoctoral researchers, and
engineers into the large U.S.\ laboratories---Fermilab, Brookhaven, and
SLAC---and, before its 1993 cancellation scattered its workforce among them, into
the Superconducting Super Collider in Texas. These were ordinary appointments held
by ordinary researchers, made because the laboratories valued the calculational
methods and instrument-building experience the \'emigr\'es carried; the
distinguished-scientist directorships were the visible peak of a
recruitment that ran far down the seniority ladder, into precisely the cohort of
capable but non-stellar scientists with which this paper is concerned.

What distinguishes this cohort from the apex is not that its members reached the
United States but that they reached \emph{independent} standing within
it---tenured and tenure-track professorships, and the permanent, continuing
staff-scientist appointments that are the national laboratories' equivalent of
tenure---rather than the service technical track of
Section~\ref{sec:ghetto-labs}. The evidence is necessarily statistical, because
the rank and file is by definition individually obscure. Where placement was
actually counted, the share is striking: an American Mathematical Society survey
found that scientists from Eastern Europe and the Soviet Union made up some fifteen
per cent of all tenured and tenure-track appointments in U.S.\ mathematics
departments in 1991--92, the overwhelming majority not medalists~\cite{borjas-doran}.
No comparable hiring census exists for the experimental sciences, but the
surrounding magnitudes agree: physical scientists were roughly half of the
former-Soviet research outflow~\cite{shiltsev-diaspora}, and some ten thousand FSU
scientists and engineers entered the United States in the
1990s~\cite{ganguli}. In the national laboratories the absorption was into the
permanent staff---the two to three thousand
former-Soviet particle and nuclear physicists entered overwhelmingly as continuing
staff scientists~\cite{shiltsev-diaspora}, in fields where roughly a third of all
permanent university and government-laboratory appointments already went to
foreign-trained doctorates~\cite{nuclear-lrp}. The collective trace is clearest at
the level of the scientific school: the Kyiv school of cellular electrophysiology
alone placed more than seventy of its alumni in professorships
abroad~\cite{kyiv-school}---a single Soviet research tradition supplying, on its
own, dozens of independent faculty appointments in the West.

Two cases treated separately below belong to this same rank-and-file absorption.
The dedicated code group of Section~\ref{sec:ghetto-labs} was an entire team of ordinary
former-Soviet scientists drawn into a single U.S.\ laboratory around a transferred
capability; and the TOPAZ program  ``eventually employ[ed] up to fifty
Russian scientists, engineers, and technicians'' in the United States---not
eminent recruits commanding chairs, but a working group paid in continued
scientific employment for the technology they brought. Taken together with the
aggregate placement evidence, these instances locate the phenomenon where the
thesis requires it: not in the brilliance of a handful at the apex, whose success
is overdetermined, but in the breadth of a demand that, for roughly a decade,
opened ordinary positions to ordinary former-Soviet scientists on terms that had
closed by the mid-2000s.

\section{The legal accelerant: the Soviet Scientists Immigration Act of 1992}

\label{act}
The premium established so far is a statement about \emph{demand}: unique Soviet
expertise was scarce and valuable, and U.S.\ institutions were willing to pay for
it in chairs, distinguished appointments, and laboratory leadership. But demand
can be realized only if the scientist can actually enter the country and be hired
without prohibitive delay---and ordinary U.S.\ employment immigration imposed
exactly such friction. A second, institutional factor of the 1990s removed that
friction for precisely this population, and so allowed the premium to be cashed
quickly. It did not create the demand; it lowered the transaction costs of
meeting it.
 
That factor was the Soviet Scientists Immigration Act of 1992, signed into law on
24~October 1992 (Public Law~102--509; 106~Stat.~3316)~\cite{ssia}.
The statute targeted scientists and engineers of the twelve former-Soviet
republics and the three Baltic states who had expertise in nuclear, chemical,
biological, or other high-technology fields, or who had worked on such
defence projects. Its two operative provisions were narrowly
practical. Section~4 directed that these scientists be designated as aliens who
``possess exceptional ability in the sciences'' for purposes of
section~203(b)(2)(A) of the Immigration and Nationality Act---the
second-preference (EB--2) employment category---\emph{whether or not} they held
advanced degrees; and Section~3 waived, for this class, the Act's requirement
that an alien's services first be sought by a U.S.\ employer~\cite{ssia}.
The authority was capped at 750 principal scientists (spouses and children
excluded) and was set to sunset four years after enactment, in October~1996; it
was later revived by the Foreign Relations Authorization Act, Fiscal Year~2003
(Public Law~107--228), with a new sunset of 30~September 2006 and a raised ceiling
of 950 visas~\cite{ssia-fedreg}.
 
The significance of these two provisions for the premium is best seen against the
default rule they displaced. An EB--2 immigrant normally needs an employer to
sponsor the petition and, in most cases, to obtain a labour certification---a slow
and costly process in which the employer must test the domestic labour market and
show that no qualified U.S.\ worker is thereby displaced. By pre-classifying these
scientists as exceptional and waiving the job-offer requirement, the Act let a
qualifying scientist pursue permanent residence \emph{without} a sponsoring offer
and \emph{without} labour certification. It thereby stripped out the single
largest source of delay and employer risk in the system: a department could
recruit a Soviet physicist without navigating certification, and the scientist
could arrive already on a path to a green card rather than on a temporary,
employer-tethered visa. In the terms of this chapter, the unique expertise
supplied the demand and the Act supplied frictionless entry; together they let the
premium be converted rapidly into academic and laboratory appointments.
 
The Act's influence, finally, extended beyond those who used its special channel
directly---which is where it bears most on the puzzle of the reversal. Even
scientists who entered by other routes (exchange and temporary-work visas, the
parallel ``extraordinary ability'' categories, or refugee and family channels)
benefited indirectly, because the Act both expressed and entrenched an official
presumption that former-Soviet scientists were exceptional and wanted, which
smoothed adjudications and signalled welcome. This is the policy complement to the
demand-side premium, and the two were temporally aligned. The special channel
lapsed in 1996, was revived only briefly between 2002 and 2006, and the broader
welcoming posture faded thereafter; as the channel narrowed and the premium itself
eroded, the rapid-placement dynamic that had defined the 1990s closed with it.
Two qualifications keep the weight of the factor in proportion. Its direct
numerical footprint was modest and capped, and in practice its evidentiary
requirements proved onerous to satisfy~\cite{ssia-fedreg}; its importance lies
therefore less in raw volume than in the frictionless channel it opened and the
posture it signalled. And immigration policy was a necessary accelerant of the
premium, not a substitute for it: it determined how quickly the premium could be
realized, but not whether the premium existed.
 
\section{The exhaustion mechanism}

The defining property of this resource was its depletability, and its mechanics explain the turn of the 2000s.

The unique knowledge was a \emph{stock, not a flow}. It had accumulated over
decades of isolated development, and the emigration of the 1990s extracted and
transferred that accumulated stock. Its carriers---including the tacit,
person-embodied components that no journal could convey---were not a thin elite
but a broad section of the Soviet research workforce; and what governed who left
first and fastest was not only scientific eminence but also the capacity and readiness to
move: foreign contacts and exposure to the West, methods and a command of language
communicable to Western colleagues, adaptability, and the willingness to seize an
opening while it lasted. The distinctive stock was thus carried out not by a
handful of the strongest, but by the large, mobile, well-connected body of ordinary
researchers best placed to move---a body selected by mobility more than by merit.
Once that stock was transferred, published, and integrated into world science, the
differential advantage dissipated: a method, having become common property, ceases
to be rare and therefore ceases to command a premium.

The bibliometric record confirms the timing. Analysis of Scopus data shows that the largest outflow of researchers from leading Russian universities occurred in the 1990s and the very early 2000s, and that it was concentrated precisely in the natural and technical sciences---the fields in which the stock of unique knowledge was greatest~\cite{scopus}. After 2000 the scale of the phenomenon became minor: those leaving for permanent residence or contract work abroad amounted to no more than about 2\% of Russian holders of higher degrees, and the very vocabulary of analysis shifted from ``brain drain'' to ``brain circulation''~\cite{intellectual-potential}. FSU ceased to be a ``black box'' of unknown contents; its scientific schools became well known, embedded in international networks, and their methods part of the common arsenal. Empirical work on the \'emigr\'e cohort itself, reconstructed from publication records, situates this exodus as a discrete historical episode rather than a standing condition~\cite{ganguli}.

The consequence for careers follows directly. In the 1990s a U.S.\ department encountering a Soviet scientist who commanded genuinely novel methods, or an entire sub-field, had a rational incentive to \emph{create} a line---and to supply start-up funding and a laboratory---because the expertise was scarce and promised to strengthen a program where the department had a gap; this is the demand that, at its apex, produced the chairs and laboratory directorships catalogued above, and survey evidence from the early wave indicates that roughly 70\% of those who left in that period found positions in universities and R\&D organizations~\cite{ird-funding}. By the mid-2000s both sources of demand had run dry: the same methods were now known and diffused, the targeted programs were winding down, and the premium for uniqueness fell to zero. A new \'emigr\'e of comparable scientific calibre entered not an environment poised to open lines for him, but an ordinary---and contracting---academic market in which he was merely one applicant among many. What had changed was not the person, but the structure of demand into which the person arrived.

Korobkov's~\cite{Korobkov2020}  examination of the Russian academic diaspora provides valuable empirical support for the demand-side exhaustion thesis advanced above. His account of the concentrated outflows from the principal scientific centres in the early 1990s, the clear predominance of basic-science fields such as physics and mathematics, and the subsequent transition to predominantly temporary migration channels with status conversion confirms that the most distinctive Soviet-era knowledge stock was transferred rapidly and that the premium it commanded could not be sustained once that stock had been assimilated into global science. Complementary survey findings coordinated by Dezhina~\cite{Dezhina2016} and her collaborators, which document that more than 97 per cent of respondents in a major 2015 diaspora sample had already resided abroad for at least a decade and that physicists continued to constitute over one-third of the flow, further illustrate the structural shift: later émigrés of comparable calibre entered an academic labour market in which formerly distinctive expertise had become widely diffused, thereby eroding the exceptional career opportunities observed during the preceding decade. Korobkov's additional observations on widespread internal brain waste and the complicating effects of post-2014 geopolitical tensions on diaspora employment prospects in security-sensitive domains complement the analysis by clarifying why the 1990s window of structural demand closed and did not reopen on equivalent terms.

This is a strong claim, and we state it with care: what was depleted was the
\emph{differential, marketable} stock---the portion of Soviet expertise that was both
distinctive and absent in the West---not the whole of Soviet science, much of which
persisted and continued to produce. That this differential stock was substantially
exhausted by the mid-2000s rests on indirect evidence---the bibliometric timing of
the outflow and the integration of Russian science into global
networks~\cite{scopus}---rather than on a direct measurement of declining uniqueness
over time, which we do not attempt and flag as a limitation.

\section{The crowded market: Asian competition and the asymmetry of ``fit''}

\label{sec:fit}
By the 2000s the market the \'emigr\'es entered had been transformed by a second,
far larger inflow---and, unlike the former-Soviet influx, the Chinese and Indian
inflow was not a one-time transfer of senior scientists but a sustained,
self-renewing graduate pipeline. International students earned more than 40\% of
the roughly half-million STEM doctorates awarded by U.S.\ universities between
2000 and 2019~\cite{cset-stay}, and the annual number of science-and-engineering
(S\&E) doctorates going to temporary-visa holders roughly doubled after 2000---up
about 101\% by 2020---peaking at 41\% of all S\&E doctorates in
2007~\cite{nsf-sed2020}. China and India have been the top two countries of origin
for U.S.\ doctorate recipients on temporary visas for at least the past decade,
together accounting for nearly half of all such doctorates, the overwhelming
majority of them in science and engineering~\cite{nsf-sed2023}. Crucially, these
graduates stayed: short-term stay rates for Chinese (83\%) and Indian (86\%) S\&E
doctorate recipients ran well above the all-country average of
73\%~\cite{nsf-stay}, and of those who earned STEM PhDs between 2000 and 2015,
roughly 90\% of Chinese and 87\% of Indian nationals were still in the United
States as of 2017~\cite{cset-stay}. The cumulative result was a large and growing
cohort of \emph{U.S.-trained} scientists competing for the same academic and
laboratory positions.
 
Against this, the former-Soviet inflow was small, mid-career, and shrinking. As
the earlier sections established, it was a one-time extraction of an existing
stock rather than a renewable flow; the bibliometric record places its peak in the
1990s and the very early 2000s, after which the emigration dwindled to no more
than about 2\% of Russian degree-holders and the literature's vocabulary shifted
from ``brain drain'' to ``brain circulation''~\cite{intellectual-potential}. The
asymmetry is visible in the doctoral pipeline itself: Russia and the other
former-Soviet states do not appear among the leading countries of origin for U.S.\
research doctorates that China and India have come to dominate~\cite{nsf-sed2023}.
The competition the FSU scientist faced in the 2000s was therefore not symmetric.
They belonged to a contracting, mid-career cohort entering a market increasingly
populated by a numerous, younger, home-trained one.  
The magnitudes can be made exact from official series: the NSF Survey of Earned
Doctorates reports doctorate recipients by country of
citizenship~\cite{nsf-sed-citizenship}, and the Department of Homeland Security
yearbooks report immigrant and nonimmigrant admissions by
country~\cite{dhs-yearbook}, against which the former-Soviet counts can be set beside
the Chinese and Indian ones for the 1990s and 2000s.

The asymmetry is not an artefact of the post-2014 estrangement; it is visible
across the friendly 1990s and 2000s in the raw counts of U.S.\ doctorate recipients
by country of citizenship. In 1998, U.S.\ universities awarded research doctorates
(all fields) to 2,571 citizens of China and 1,259 citizens of India, but to only
216 citizens of Russia~\cite{sed1998}; by 2005 the figures were 3,827 for China,
1,274 for India, and 255 for Russia~\cite{sed2005}.  
The gap is not peculiar to Russia; it characterises the former Soviet space as a
whole. None of the other fourteen republics appears in
either year---neither Ukraine, Belarus, nor Moldova; nor the Baltic states (Estonia,
Latvia, Lithuania); nor the South Caucasus (Georgia, Armenia, Azerbaijan); nor the
Central Asian states (Kazakhstan, Kyrgyzstan, Tajikistan, Turkmenistan,
Uzbekistan)---which means each of them produced no more than the thirtieth-ranked
total of 66 recipients in 1998 and 76 in 2005~\cite{sed1998,sed2005}. Russia is thus
the dominant component of any former-Soviet aggregate, and the non-Russian remainder,
composed entirely of countries individually below those cutoffs, is modest. 

Moreover, even this limited outflow from Russia was highly concentrated among graduates of a small number of elite institutions. Available observations and bibliometric patterns indicate that students from Lomonosov Moscow State University (MSU) and the Moscow Institute of Physics and Technology (MIPT) accounted for the large majority of those who pursued doctoral studies in the United States. Dedicated bridging programs that facilitated such transitions were rare and small in scale. One example is the PARTI (Physics of Accelerators and Related Technology for International Students) summer internship program at Fermilab’s Accelerator Physics Center, which enabled selected physics and engineering students from the former Soviet Union to conduct research at the laboratory; some participants subsequently obtained longer-term positions or entered U.S.\ Ph.D.\ programs. Such initiatives typically trained only a few dozen students per year across participating countries. Based on my observations, MIPT and Novosibirsk State University were especially well represented among PARTI participants. The PARTI program was discontinued after the 2014 annexation of Crimea as institutional collaborations with Russian universities were curtailed \cite{PARTI-Fermilab}.

An exact
post-Soviet total is not given in the published ranking and would require a custom
tabulation of the SED microdata by country of citizenship, available through NCSES's
WebCASPAR and interactive data tools~\cite{ncses-sed-tool}; but even a generous
allowance---crediting the fourteen other republics with a combined one to two hundred
recipients on top of Russia's---leaves the entire former USSR producing only a few
hundred U.S.\ doctorate recipients a year. The
demographic asymmetry behind the competition described here is therefore a feature of
the whole former-Soviet scientific diaspora, not of Russia in isolation.
 Over those seven years Chinese
output grew by roughly half (a factor of 1.49) and Russian output by less than a
fifth (1.18), so that the China-to-Russia ratio widened from about 12:1 to about
15:1, while India alone produced five to six times as many U.S.\ doctorate
recipients as Russia throughout.

The contrast in magnitude is stark. In
the pooled 2017--2019 cohort of S\&E doctorate recipients on temporary visas China
supplied 36\% (about 15,800) and India 13\% (about 5,750), whereas all of Europe
excluding Turkey---of which Russia is only a part---supplied 7\%~\cite{nsf-stay}.
A self-renewing pipeline numbered in the thousands per year on one side, and a few
score on the other, is the demographic fact underlying the asymmetric competition
described here.
 
That difference in entry mode---mid-career arrival versus a doctorate earned inside
the U.S.\ system---translated into a difference in ``fit'' that the U.S.\
tenure-track market rewards heavily, and the point is structural rather than a
matter of aptitude. The market is steep and narrow: the production of S\&E
doctorates has far outrun the supply of faculty lines~\cite{pnas-nexus}, so that by
one estimate only about 13\% of S\&E PhD graduates obtain a tenure-track
position~\cite{pmc-anthro}; and it is a prestige hierarchy in which roughly 80\% of
tenure-track faculty are trained at just 20\% of U.S.\
universities~\cite{nature-hiring}. Success in it depends on fluency in tacit
rules---publication strategy, grant-writing in the idiom of U.S.\ funding agencies,
and the assembly of a near-tenure-ready record---and on embeddedness in the
mentoring and recommendation networks through which appointments are in practice
made~\cite{pmc-anthro}. 

These competences are acquired chiefly through U.S.\
doctoral and postdoctoral training. Chinese and Indian competitors, having passed through that training, arrived already fluent in the rules and embedded in the
networks; the FSU scientist, formed in a different system and arriving mid-career,
did not, and faced the added task of learning the game while competing in it. In
the 1990s the expertise premium had overridden this disadvantage---departments
created positions for unique capabilities irrespective of how well their holders
fitted the local academic culture. By the 2000s the FSU scientist had lost the
premium and faced, at the same time, a numerous and
better-fitted competition. The two changes were mutually reinforcing: the scarcity
that had once made fit irrelevant was gone, and the field was now crowded with
candidates for whom fit was, in effect, second nature.

\begin{table}[ht]
\centering
\small
\begin{tabular}{lcc}
\hline
 & China & India \\
\hline
Among top-two origins, U.S.\ temporary-visa doctorates (past decade) & yes & yes \\
Short-term stay rate, S\&E doctorate recipients & 83\% & 86\% \\
Long-term stay (2000--2015 STEM PhDs, present in U.S.\ in 2017) & $\sim$90\% & $\sim$87\% \\
\hline
\end{tabular}
\caption{The Chinese and Indian doctoral pipeline and its retention (all-country
short-term stay average: 73\%). Sources:~\cite{nsf-sed2023,nsf-stay,cset-stay}.}
\label{tab:asian-pipeline}
\end{table}

\section{The post-9/11 turn: a headwind for the foreign-born at large}
 
The fourth factor differs from the first three in that it was not specific to the
former-Soviet cohort, or indeed to any one group: the attacks of 11~September 2001
reset the terms of U.S.\ immigration broadly, and their effects reached the
foreign-born at large. The most general of these was a change of frame. Before
9/11 the American immigration debate had been conducted largely in the language of
labour and the economy; afterward it was conducted in the language of national
security, and that securitization attached to immigration as a whole rather than to
a single category of immigrant~\cite{uva-leblang}. Panel-survey evidence indicates
that natives' attitudes toward immigration in general turned more restrictive in
the weeks following the attacks, although that particular shift proved
short-lived~\cite{socialforces-skin}; the more lasting consequence was the energy
the moment gave to nativist and restrictionist politics that opposed immigration of
every kind~\cite{uva-leblang}, and the climate of official scrutiny---including
round-ups of the foreign-born---that unsettled immigrant communities well beyond
those directly suspected~\cite{globalboston}. It is important to be precise about
distribution: the acute hostility of the period, including violence and profiling,
fell overwhelmingly on Muslim, Arab, and South Asian communities, and on those
mistaken for them~\cite{globalboston}. The claim here is the complementary one---that
the institutional regime and the general climate formed a headwind for foreign
scientists of every origin, the former-Soviet included.
 
That institutional regime is the most concrete and the most clearly broad of the
mechanisms, because it operated on scientific fields and visa categories rather
than on ethnicity. After 9/11 the Student and Exchange Visitor Information System
(SEVIS), a tracking database for the more than half a million international
students, scholars, and scientists in the country, was fast-tracked into
operation, and almost all visa applicants were newly required to appear for an
in-person interview~\cite{physicstoday-visa,gao-higher-ed}. Two security-screening
programmes bore directly on scientists. Visas Mantis, a clearance required of
foreign students and scholars working in any of roughly two hundred designated
scientific and technical fields, applied irrespective of nationality; Visas Condor,
added in 2002, flagged nationals of designated countries of concern for additional
review~\cite{gao-mantis,physicstoday-visa}. The agencies were unprepared for the
workload, and the result was a large backlog and waits that could run to months,
with applicants in sensitive fields particularly exposed~\cite{nas-visa,physicstoday-visa}.
The deterrent effect was acknowledged at the time: the State Department conceded
that long waits could discourage legitimate travel and damage foreign citizens'
impressions of the country~\cite{gao-higher-ed}; the Association of American
Universities reported that the clearances had inconvenienced thousands of
international students and discouraged many more from coming~\cite{gao-higher-ed};
and the European scientific press warned that the new barriers were turning
researchers away~\cite{nas-visa}. The aggregate is visible in the enrolment record:
universities reported that visa denials and delays were the single largest cause of
declines in new international students in this period~\cite{wenr-recruit}, and the
post-9/11 years stand as one of only two interruptions---the COVID-19 pandemic being
the other---to the long-run growth of international enrolment before
2020~\cite{highered-dive}. Although the backlogs were substantially cleared by late
2004~\cite{nas-visa}, the screening and tracking infrastructure became permanent.
 
For the question of this chapter the regime's bearing is twofold. For
former-Soviet scientists already settled in the United States---most of the cohort
that concerns us---the visa mechanics mattered less than the surrounding climate,
since they were already in place; what they experienced was the diffuse chill of a
society that had turned more security-minded toward foreignness, a chill consistent
with the documented shift even where its day-to-day, interpersonal expression (cooler
collegial relations, narrowed informal networks) is harder to document directly than
the policy record. For new arrivals, and for the cohort as a whole, there was in
addition a sharper and more specific bite that connects back to the premium itself:
the fields in which Soviet expertise had been most uniquely valuable---nuclear,
chemical, and biological science---were precisely the security-sensitive fields now
subject to the heaviest screening and suspicion. The securitization of 2001 thus
worked, in part, to convert the very expertise that had been a premium asset in the
1990s into a marker of risk in the 2000s.  
It should be added, for the former-Soviet cohort specifically, that the dedicated
immigration channel of the 1992 Act was already lapsing by 2001 (Section~\ref{act});
9/11's marginal effect on this group therefore operated less through the closure of a
special programme---which was ending in any case---than through the general climate
and through the heightened vetting that fell on their security-sensitive fields.

One qualification remains, and it cuts the other way: the weight of this factor
should not be understated. Granted, the measured shift in public attitudes toward
immigration, real as it was at the time, proved partly
transient~\cite{socialforces-skin}, and the acute hostility of the period fell most
heavily on particular communities rather than on the immigrant-origin population at
large~\cite{globalboston}. But neither observation touches the mechanism that
mattered for this cohort. What endured was not a poll figure but a colder climate
toward the non-native in academic and laboratory life, and its damage ran through
the informal channel on which advancement actually turns. Tenure and promotion are
settled not by publications alone but through letters of recommendation, sponsorship
by senior colleagues, inclusion in the coalitions that carry a case through a
department, and the collaborative ties that build a reputation---exactly the
trust-based relationships that a securitized wariness of the outsider quietly
corrodes. And this told against the former-Soviet cohort where it was already
weakest: arriving mid-career without the networks a U.S.\ doctorate confers
(Section~\ref{sec:fit}), and frequently placed in co-national enclaves only loosely
joined to the mainstream (Section~\ref{sec:ghetto-labs}), these scientists depended
on forging precisely the cross-group relationships that the post-2001 chill made
harder to forge---and the chill arrived just as the expertise premium that had once
opened doors regardless of such frictions was itself eroding. Its durable form was
institutional and atmospheric rather than a matter of settled opinion, which made it
quieter than the acute episodes but no less consequential for who was promoted, who
was sponsored, and who advanced to independent standing.

\section{Conclusion}
 
This paper set out to explain a puzzle that the dominant narrative of the
post-Soviet scientific exodus does not adequately resolve: why scientists who
reached the United States in the early and middle 1990s so often secured
prestigious and visible academic positions, while those of comparable calibre who
arrived after about 2000 built markedly more modest, less visible, and frequently
non-academic careers. Declining to treat the question as one about the thin apex of
Nobel- and Fields-level émigrés---whose early departure and brilliant ascent explain
little and risk tautology---we fixed attention on the far larger cohort of capable
but non-stellar scientists, and asked why the same kind of person fared so
differently according to the year of arrival. That framing forces the explanation
away from the attributes of the people and toward the structure of the market that
received them~\cite{graham-dezhina}, and we examined four conditions that were
unusually favourable in the 1990s and that closed, more or less together, by the
middle of the 2000s: the technology transfer and export of unique Soviet expertise;
the favourable U.S.\ immigration regime created by the Soviet Scientists Immigration
Act of 1992; the drastic enlargement of the Chinese and Indian doctoral inflow; and
the aftermath of 11~September 2001.
 
On the balance of the evidence, all these four were materially in
play. Technology transfer and export, the immigration regime of 1992, and the Asian
competition each left a documentable mark on the careers of the cohort. the consequences of 9/11 were the most diffuse of the four but neither the
slightest nor merely a passing mood. The measured spike in restrictive sentiment
faded~\cite{socialforces-skin}, yet the securitized wariness of the non-native that
outlasted it worked through the very machinery on which advancement
turns---the recommendations, senior sponsorship, and departmental coalitions behind
any grant of tenure, promotion, or an independent line---and it told hardest on a
cohort already poorest in those ties: mid-career, network-thin, and often enclaved,
meeting this chill just as the expertise premium that had once overridden such
frictions was giving way.

Among the factors that were in play, technology transfer and export was
primary. It was the engine of the whole episode: the distinctive,
isolated-developed expertise carried out of the USSR by the first cohorts was a
finite national stock that commanded a genuine premium in the American research
market, and that premium---paid in chairs, laboratory leadership, and positions
created to absorb it---is what made the 1990s a window. The other factors are best
understood in relation to it. The 1992 Act did not create the premium but lowered
the friction of realizing it, letting it be cashed quickly into appointments before
the Act lapsed; the Asian competition mattered most once the premium was gone,
filling on ordinary terms the positions that scarcity had previously reserved; and
the securitization that followed 9/11 bore, with particular irony, on the very
weapons-relevant fields in which Soviet expertise had been most valuable. When the
stock was extracted, published, and absorbed into global science by the mid-2000s,
the premium collapsed, and with it the surplus demand on which the favourable window
had rested. This is why the depletion of an exported, finite resource---rather than
any decline in the abilities of successive émigrés---is the primary explanation of
the reversal.
 
A further mechanism, cutting across the four, compounded the difficulty and deserves
emphasis in its own right: cultural ghettoization. Even within the favourable
window, and especially in the laboratory setting, a portion of the émigré cohort was
absorbed not into independent careers but into co-national enclaves---groups led by
more successful compatriots and staffed with Russian-speaking subordinates---in
which, whatever the goodwill of those who led them, the structural arrangements of
the host institution left their members with limited scientific visibility, weak
integration into the mainstream networks through which academic careers are made, and
few routes to independent standing~\cite{wagner}. This enclave dynamic was a genuine
channel into employment, but it was also a ceiling; and as the favourable conditions
of the 1990s gave way, it increasingly defined the kind of position---subordinate
rather than independent---that a later arrival could expect. Ghettoization thus
operated as an additional, internal brake on the careers of the cohort, reinforcing
from within the constraints that the four external factors imposed from without.
 
The bottom line is that the imbalance between the émigré generations was structural
rather than personal. The cohort that arrived first did not, for the most part,
possess extra abilities that the later cohort lacked; it arrived into a configuration
of demand, policy, competition, and climate that the later cohort did not enjoy, and
it found in that configuration opportunities that had largely closed a decade later.
The comfortable story in which the cream left first and the rest were ordinary
mistakes the shape of a market for the shape of a people. What changed, above all,
was not who was leaving FSU, but what awaited them when they arrived---and the factor we judge primary among several that changed was the exhaustion of the exported,
depletable technology and expertise for which, in the 1990s, premium careers had been the price.

A final point bears on fairness as much as on explanation. In attributing the career
gap to the changing composition of the flow---the elite first, the ordinary
after---the received account~\cite{graham-dezhina} did more than misdescribe a
labour market; however reasonable it appeared, it lent scholarly weight to the
impression that each successive wave of former-Soviet \'emigr\'es was of lesser
quality than the one before, an impression the evidence assembled here does not bear
out. Once in circulation, such a perception is not inert: it can quietly shadow later
arrivals in precisely the settings---search committees, collaborations, letters,
promotion cases---where reputations are weighed and careers made. If the reversal was
structural rather than personal, then those who came later were not the diminished
residue of a departed elite but scientists of comparable calibre who met a less
favourable market, and fairness to them begins with saying so plainly.

\end{document}